\documentclass[a4paper, 10pt, conference]{ieeeconf}      % Use this line for a4 paper

\IEEEoverridecommandlockouts                              % This command is only needed if 
\usepackage{authblk}
\usepackage{amsfonts}
\usepackage{amsmath}         
\usepackage{hyperref,cite}
\usepackage{cleveref}
\usepackage{color}
\usepackage{algorithm}
\usepackage{algorithmic}
\let\labelindent\relax
\usepackage{enumitem}
\usepackage{float}
\usepackage{graphicx}
\usepackage{booktabs}
\usepackage{threeparttable}
\usepackage{subfigure}

\begin{document}
\title{\LARGE \bf
An interpretative and adaptive MPC for nonlinear systems}
% \author{Liang Wu$^{1}$, Alberto Bemporad$^{1}$%
% %\thanks{*This work was not supported by any organization}% <-this % stops a space
% \thanks{The authors are with the IMT School for Advanced Studies Lucca, Italy,
%         {\tt\small \{liang.wu,alberto.bemporad\}@imtlucca.it}}%
% }
\author{Liang Wu$^{1}$%
%\thanks{*This work was not supported by any organization}% <-this % stops a space
\thanks{The authors are with the IMT School for Advanced Studies Lucca, Italy,
        {\tt\small liang.wu@imtlucca.it}}%
}
\thispagestyle{empty}
\pagestyle{empty}
\maketitle

\begin{abstract}
Model predictive control (MPC) for nonlinear systems suffers a trade-off between the model accuracy and real-time computational burden. One widely used approximation method is the successive linearization MPC (SL-MPC) with EKF method, in which the EKF algorithm is to handle unmeasured disturbances and unavailable full states information. Inspired by this, an interpretative and adaptive MPC (IA-MPC) method, is presented in this paper. In our IA-MPC method, a linear state-space model is firstly obtained by performing the linearization of a first-principle-based model at the initial point, and then this linear state-space model is transformed into an equivalent ARX model. This interpretative ARX model is then updated online by the EKF algorithm, which is modified as a decoupled one without matrix-inverse operator. The corresponding ARX-based MPC problem are solved by our previous construction-free, matrix-free and library-free CDAL-ARX algorithm. This simple library-free C-code implementation would significantly reduce the difficulty in deploying nonlinear MPC on embedded platforms. The performance of the IA-MPC method is tested against the nonlinear MPC with EKF and SL-MPC with EKF method in four typical nonlinear benchmark examples, which show the effectiveness of our IA-MPC method.
\end{abstract}


\section{Introduction}\label{sec:intro}
Model Predictive Control (MPC) is an advanced technique to control multi-input multi-output systems subject to constraints~\cite{garcia1989model}. MPC has been widely used in diverse industrial areas, such as process \cite{qin2003survey}, aerospace \cite{eren2017model}, power electronics \cite{geyer2016model}, etc. 
The core idea of MPC is to predict the evolution of the controlled system through a dynamical model, solve an optimization problem over a finite time horizon, only implement the control input at the current time, and then repeat the optimization at the next sample step~\cite{qin2003survey,bemporad1999robust}. 

After major developments in the field of MPC over the past three decades, MPC for linear plants described by the linear state-space model has made significant progress in theoretical stability analysis and real-time numerical algorithm implementation \cite{lee2011model}. However, most industrial plants are nonlinear, and their constrained nonliner MPC (NMPC) formulation is a non-convex optimization problem that encounters practical difficulties in terms of computational complexity and algorithm implementation, such as solving it within sampling time on embedded platforms. It's known that a trade-off exists between the described accuracy of the nonlinear model and the online NMPC computation cost: the more accurate/complicated the model, the greater the online computational cost.

\subsection{Related works}\label{sec: related}
Most industrial approaches are based on linear models obtained from the linearization technique in solving MPC problems for nonlinear systems. A nonlinear plant generally admits a locally-linearized model when considering regulating a particular operating point. In most cases, a linear MPC yield adequate performance, especially for chemical process industries. This approach with one linear model has an advantage in designing and deploying the offline explicit MPC solutions based on multi-parametric quadratic programming \cite{bemporad2002explicit} and the online MPC solutions based on fast convex optimization algorithms for embedded platforms to satisfy real-time requirements. For tracking problems that operate over a wide range of operating conditions, multiple linear model based MPC addresses the non-adequate accuracy issue in one linear model based MPC. Multiple linear models are also called linear parameter varying (LPV), often used in the context of gain-scheduling, namely an MPC controller scheduled by linear models \cite{chisci2003gain}. One approach for obtaining an LPV model is the online successive linearization at the current states based on a first-principles model \cite{kuhne2004model}. 

One improved approach is the linear time-varying (LTV) model, obtained from linearizing the nonlinear model at each prediction horizon \cite{falcone2008linear}. However, the LTV-MPC approach comes at a higher computational cost than the LPV-MPC approach. Since the states and inputs during the prediction horizon are unknown, the typical way utilizes the shifted optimal inputs sequence of the previous MPC solution as the inputs, and the states are calculated by integrating the nonlinear model. Another successful and broadly used approach is the real-time iteration (RTI) scheme. The RTI approach assumes that the MPC solution at the current time is very similar to the solution obtained at the previous time. Under this assumption, a full Newton step can be taken, providing an excellent approximation of the fully converged NMPC solution. Not only that, but the RTI algorithm also divides the whole computation into preparation phase and feedback phase phases, to achieve a shorter feedback delay \cite{gros2020linear}. The RTI-NMPC approach has been implemented into the open-source software ACADO Toolkit, which allows exporting the optimized C-code for deployment \cite{quirynen2015autogenerating,houska2011auto}. Another approximate scheme in the literature is the Continuation/GMRES method \cite{ohtsuka2004continuation}. In \cite{ohtsuka2004continuation, ohtsuka2015tutorial,katayama2020automatic}, the authors also provide its full implementation within the tool AutoGenU to support the C-code generation.

In addition to the above approximating approaches, incorporating nonlinearity directly into the MPC problem can provide a systematic way of dealing with systems with nonlinear dynamics, constraints, and objectives but at the cost of heavy online computation. Solving the resulting non-convex optimization problem relies on an efficient and reliable nonlinear programming algorithm, which has been alleviated to some extent with the advent of tailored and professional software tools, such as the CasADi \cite{Andersson2019} and FORCES NLP \cite{zanelli2020forces}.

To further reduce the online computation cost of NMPC, the tremendous advances in machine learning and deep learning inspired many works, such as learning a globally-linear model or learning an efficient representation of approximated MPC laws via deep neural networks. Herein we only name a few works related to MPC for nonlinear systems. In \cite{korda2018linear}, the author utilized a lifting operator, called the Koopman operator, to lift the nonlinear dynamics into a higher dimensional space where its evolution is approximately linear. Then an MPC controller based on the lifted linear model can replace the NMPC. The lifted linear model can be learned by a user-defined dictionary library or deep neural networks from data \cite{lusch2018deep,kaiser2021data}. Another approach that directly learns the approximated MPC law to reduce the online computation cost has been explored in the literature. In \cite{karg2020efficient}, the authors show that a neural network can represent exactly the MPC law described by the piecewise affine function in the case of linear MPC. The authors further extended to the case of nonlinear systems \cite{lucia2018deep}, in which the deep neural network is utilized to learn the robust nonlinear MPC law.

% 讲述EKF
In the practice of the MPC technique, in addition to numerical optimization algorithms needed to solve the MPC problem, another critical ingredient is estimation algorithms. Generally, the nonlinear dynamic model can be an interpretative first-principle state-space model or a black-box model built from experimental data via system identification (or modern machine learning) in an NMPC setting. A nonlinear first-principle state-space model can provide interpretability that is preferred in practice, but it often suffers the issues like unavailable full-state information, unmeasured disturbance, or unmodelled time-varying terms. Therefore, these cases require a state estimation algorithm or a joint state parameter estimation algorithm in which unmeasured disturbances and unmodeled time-varying terms are considered as parameters to be estimated. Many state estimation methods for nonlinear systems exist. The well-known extended Kalman Filter (EKF) is undoubtedly the predominant state estimation technique \cite{norgaard2000new}, and the EKF has also been applied in many joint state-parameter estimation applications \cite{wan2001dual,fux2014ekf}.

The identified nonlinear black-box model used in NMPC can be divided into two categories: state-space or input-output formulation, such as state-space based recurrent neural networks (RNNs) or input-output based neural-network autoregressive model with exogenous inputs (NNARX). One obvious advantage of input-output based models over state-space based models is that input-output based models don't require an state estimation algorithm to estimate the generally higher-dimensional states. Despite this, input-output based models still require an online estimation algorithm to update the model parameters when discrepancies happen between the model output predictions and the streaming output measurements. The linear input-output ARX models are well known for their adaptability allowing efficient online recursive estimation algorithms, like recursive least-squares (RLS) or Kalman Filter \cite{guo1995performance}. In fact, nonlinear ARX models also have good adaptability, updating their model parameters online via the EKF algorithm \cite{bemporad2021recurrent}.

As discussed above, the EKF algorithm is an essential component in NMPC practice, both for the first-principle based and black-box models. The EKF is based on the first-order linearization of nonlinear dynamics around the previously estimated vectors. Although the first-order linear approximation used by the EKF is sometimes not accurate enough, other advanced techniques can provide better estimation accuracy, but this comes at the cost of complicated implementation and expensive computational costs. The ease of implementation and computation burden are of great importance, especially in the cooperation of NMPC and EKF. Despite the widely practical usefulness of the EKF, its convergence and stability guarantee requires strict conditions, such as satisfying the nonlinear observability rank condition and having sufficiently small initial estimation error as well as disturbing noise term \cite{reif1999stochastic,ljung1979asymptotic}. In this paper we assume that the EKF algorithm will not suffer from divergence problems in its applications.

It is worth noting that the widely used SL-MPC approach and the EKF algorithm are both based on linearization techniques. Their combination, namely the SL-MPC with the EKF method presented in \cite{lee1994extended}, has become a common choice to handle the nonlinear first-principle based model in the industrial practice of NMPC technique.In its EKF part, the states and unknown disturbances are recursively updated from the streaming input-output data based on the linearization of the discrete-time nonlinear model. In its SL-MPC part, the control input is calculated based on the estimated states and disturbances at the current time and the linearized state-space model, which is also updated online based on the linearization of a first-principle model at the estimated states and disturbances. Thus, the SL-MPC with the EKF method, in a sense, is utilizing the EKF algorithm and the streaming input-output data to continuously update online a linear state-space model, which is under the representation of states defined by the first-principle-based model. And a linear input-output plant could have infinite equivalent linear state-space realizations via coordinate transformation. Our previous work \cite{wu2022equivalence} also illustrated that a linear state-space model can be equivalently transformed into an input-output ARX model, which is well known for its adaptibility. In this paper, the SL-MPC with EKF method is the starting point to derive our interpretative and adaptive MPC (IA-MPC) method, which is illustrated in detail in Section \ref{sec:ekf_nmpc}.

\subsection{Contribution}\label{sec:contribution}
In our IA-MPC method, a linear state-space model is first obtained by performing the linearization of a first-principle based model at the initial point, then transformed into an equivalent ARX model; all of these are performed offline. This offline ARX model acquisition not only makes the black-box ARX model inherit the interpretability of the first-principle based model but also exploits its adaptive nature; this is why we call our method Interpretive and Adaptive MPC (IA-MPC). In the online closed-loop control, the EKF algorithm is used to update the ARX model parameters, and our previous construction-free CDAL-ARX algorithm \cite{wu2022construction} is used to compute the MPC control input. 

The advantage of our IA-MPC method are sumarized as follows:
\begin{enumerate}
\item In addition to the gained interpretability and adaptability, the acquisition of the offline ARX model also makes it possible to have a sufficiently small initial estimation error in ARX model parameters. All the EKF algorithm needs to do is the system tracking (or feedback correction), not the system identification.
\item Compared to the EKF algorithm in the SL-MPC method, the EKF algorithm in our IA-MPC method can be decoupled when updating the ARX model parameters, thanks to the decoupling feature of ARX model. The decoupled EKF computation avoids matrix inversion and only involves scalar division, even allowing parallel calculation according to the number of outputs.
\item The corresponding ARX-based MPC problem is solved by our previous construction-free, matrix-free, and library-free CDAL-ARX algorithm. Thus,  our IA-MPC method would significantly reduce the difficulty in deploying nonlinear MPC on embedded platforms.
\end{enumerate}

\section{Successive Linearization NMPC with EKF}\label{sec:ekf_nmpc}
\begin{figure}
\centering
        \hspace*{-1em}\includegraphics[width=0.8\columnwidth]{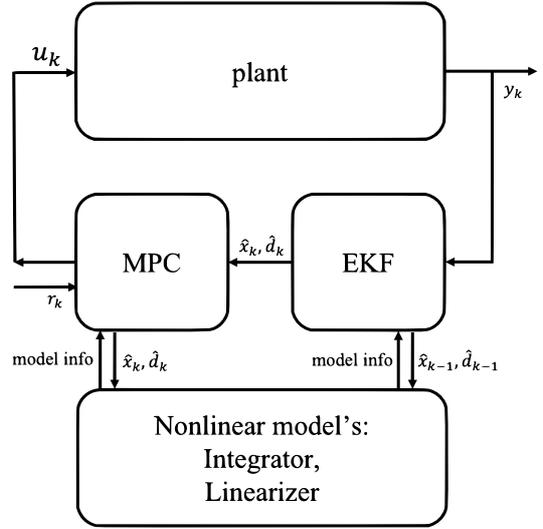} 
        \caption{Schematic diagram of successive linearization NMPC with EKF}
        \label{EKF_MPC} 
\end{figure}

This section considers a MPC tracking problem for nonlinear systems subject to the input and output constraints, considering a first-principle nonlinear model as follows
\begin{subequations}\label{sys0}
\begin{eqnarray}
&\dot{x} = f(x,u,d)\label{sys0_f} \\
&y = g(x,d)\label{sys0_g}
\end{eqnarray}
\end{subequations}
where $x \in \mathbb{R}^{n_x}$, $u \in \mathbb{R}^{n_u}$, and $y \in \mathbb{R}^{n_y}$ are the states, inputs, and outputs, respectively. And $d \in \mathbb{R}^{n_d}$ denotes the unmeasured disturbances. In industrial NMPC practice, one efficient approach for handling the nonlinearity of a first-principle model is based on successive linearization, which is applied both in state estimation and MPC computation \cite{lee1994extended}. Then, an output-feedback nonlinear MPC combines a state estimator (EKF) and a state-feedback successive linearization NMPC, and its schematic diagram is shown in Fig. \ref{EKF_MPC}.

\subsection{Extended Kalman Filter}\label{sec:ekf}
The extended Kalman Filter exploits the idea that performs the linearization at each sampling time to approximate the nonlinear system as a time-varying system and apply the linear filtering theory. In digital controller design, $u$ and $d$ are assumed to a constant value between the sampling time. Thus, a discrete-time model of (\ref{sys0}) can be formulated as follows:
\begin{subequations}\label{line_sys_ekf}
\begin{eqnarray}
&x_{k} = F_{t_s}(x_{k-1},u_{k-1},d_{k-1})\label{line_sys_ekf_F} \\
&y_k = g(x_{k},d_{k})
\end{eqnarray}
\end{subequations}
where $F_{t_s}(x_{k-1},u_{k-1},d_{k-1})$ represents the the terminal state vector obtained by integrating the ordinary differential equation (\ref{sys0_f}) for one sample interval $t_s$ with the initial condition of $x_{k-1}$ and constant inputs of $u_{k-1}$ and $d_{k-1}$. For a better state estimation, the unmeasured disturbance should also be estimated simultaneously. In general, $d$ can be modeled as the following stochastic difference equation
\begin{equation}\label{eqn_disturbance}
    d_{k} = d_{k-1} + w_{k-1}
\end{equation}
where $w_{k-1}$ is white noise sequence.
Combining the Eqn (\ref{line_sys_ekf}) with (\ref{eqn_disturbance}), the following augmented model is obtained
\begin{subequations}\label{sys_ekf}
\begin{eqnarray}
\left[\begin{array}{c}
     x_k \\
     d_k 
\end{array}\right]&=&\left[\begin{array}{c}
     F_{t_s}(x_{k-1},u_{k-1},d_{k-1}) \\
     d_{k-1} + w_{k-1}
\end{array}\right]\label{sys_ekf_prediction}\\
y_{k} &=& g(x_{k},d_{k}) + v_{k}
\end{eqnarray}
\end{subequations}
where the measurement noise sequence $v_{k}$ is often to appear in the measured output $y_k$. By linearizing  (\ref{sys_ekf}) at $\{\hat{x}_{k-1},\hat{d}_{k-1}\}$, the EKF computes the new estimates $[\hat{x}_{k}^{\prime},\hat{d}_{k}^{\prime}]^{\prime}$ from the feedback measurement $y_k$ and the model prediction (\ref{sys_ekf_prediction})
\begin{subequations}
\begin{eqnarray}
\left[\begin{array}{l}
x_{k} \\
d_{k}
\end{array}\right] &\approx& \left[\begin{array}{l}
F_{t_s}\left(\hat{x}_{k-1}, u_{k-1}, \hat{d}_{k-1}\right) \\
\hat{d}_{k-1}
\end{array}\right] \\
&+&\Phi_{k-1}\left[\begin{array}{l}
x_{k-1}-\hat{x}_{k-1} \\
d_{k-1}-\hat{d}_{k-1}
\end{array}\right]+\left[\begin{array}{l}
0 \\
I
\end{array}\right] w_{k-1}\nonumber\\
 y_k &\approx& g\left(\hat{x}_{k-1},\hat{d}_{k-1}\right)\\
&+&\Theta_{k-1}\left[\begin{array}{l}\nonumber
x_{k-1}-\hat{x}_{k-1} \\
d_{k-1}-\hat{d}_{k-1}
\end{array}\right]+v_k
\end{eqnarray}
\end{subequations}
where
\begin{equation}
\begin{aligned}
\Phi_{k-1}=\left[\begin{array}{cc}
\Lambda_{k-1} & \Lambda_{k-1}^d \\
0 & I
\end{array}\right]\\
\Theta_{k-1}=\left[\begin{array}{cc}
    H_{k-1} & H_{k-1}^d
\end{array}\right]    
\end{aligned}
\end{equation}
$\Lambda_{k-1}$,$\Lambda_{k-1}^d$, $H_{k-1}$ and $H_{k-1}^d$ are calculated using the following formula
\begin{subequations}
\begin{eqnarray}
&\Lambda_{k-1}=\frac{\partial F_{t_s}(x, u, d)}{\partial x} \mid_{x=\hat{x}_{k-1}, u=u_{k-1}, d=\hat{d}_{k-1}}\\
&\Lambda_{k-1}^d=\frac{\partial F_{t_s}(x, u, d)}{\partial d} \mid_{x=\hat{x}_{k-1}, u=u_{k-1}, d=\hat{d}_{k-1}}\\
&H_{k-1}=\frac{\partial g(x,d)}{\partial d}\mid_{x=\hat{x}_{k-1}, d=\hat{d}_{k-1}}\\
&H_{k-1}^d=\frac{\partial g(x,d)}{\partial d}\mid_{x=\hat{x}_{k-1}, d=\hat{d}_{k-1}}
\end{eqnarray}
\end{subequations}
where $\partial F_{t_s} / \partial x$ and $\partial F_{t_s} / \partial d$ represent Jacobian matrices of $F_{t_s}$ with respect to $x$ and $d$, respectively. And $\partial g / \partial x$ and $\partial g / \partial d$ represent Jacobian matrices of $g$ with respect to $x$ and $d$, respectively. Thus, we can conclude the EKF computation procedure which has been well presented and analysed, see Tab. \ref{tab:EKF-NMPC}.

\begin{table}[!htbp]
    \caption{EKF for state estimation}
    \centering
    \begin{tabular}{ll}
    \hline\hline% \toprule
        Step &  Formula\\
    \midrule
        Initialization & $P_0=\epsilon I$, $\epsilon$ is a large number,\\
        &  Initial guess $\hat{x}_0,\hat{d}_0$,\\
        & $Q$ is the process noise covariance of $\{x,d\}$,\\
        & $R$ is the measurement noise covariance.\\
        \midrule
        1-Prediction &
        $\left[\begin{array}{c}
            \hat{x}_k  \\
            \hat{d}_k 
        \end{array}\right]=\left[\begin{array}{c}
            F_{t_s}(\hat{x}_{k-1},u_{k-1},\hat{d}_{k-1}) \\
            \hat{d}_{k-1}
        \end{array}\right]$\\
        2-Correction &
        $\left[\begin{array}{c}
            \hat{x}_k  \\
            \hat{d}_k 
        \end{array}\right]=\left[\begin{array}{c}
            \hat{x}_k^{-}  \\
            \hat{d}_k^{-} 
        \end{array}\right]+K_k\left(y_k-g(\hat{x}_k^{-},\hat{d}_k^{-})\right)$
        \\
        \midrule
        \multicolumn{2}{c}{$P_k^{-}=\Phi_{k-1}P_{k-1}\Phi_{k-1}^{\prime}+Q$}\\
        \multicolumn{2}{c}{$K_k=P_{k}^{-}\Theta_{k-1}^{\prime}\left(\Theta_{k-1}P_k^{-}\Theta_{k-1}^{\prime}+R\right)^{-1}$}\\
        \multicolumn{2}{c}{$P_k=\left(I-K_k\Theta_{k-1}\right)P_k^{-}$}\\
    \hline\hline% \bottomrule    
    \end{tabular}
    \label{tab:EKF-NMPC}
\end{table}

\subsection{Successive Linearization NMPC}\label{sec:sl_nmpc}
By recursively handling the streaming input-output data $(u_{k-1},y_{k-1})$, the EKF algorithm could calculate the new estimated states and disturbances $(\hat{x}_{k},\hat{d}_{k})$. The $(\hat{x}_{k},\hat{d}_{k})$ is not only used as the nominal values to linearize the nonlinear first-principle model but also used as the initial conditions of the linearized state-space model in the SL-MPC method. Note that the continuous-time model~(\ref{sys0}) has to be transformed into the discrete-time model for digital MPC design. Two approaches, \textit{first discretise then linearise} and \textit{first linearise then discretise}, have been reported. The EKF algorithm presented in Section \ref{sec:ekf} is based on the \textit{first discretise then linearise} approach. In the context of SL-MPC, it is more common to use the \textit{first linearise then discretise}. Here the continuous-time first-principle-based model (\ref{sys0}) is linearized at current points $\{\hat{x}_k,u_{k-1},\hat{d}_k\}$,

\begin{subequations}\label{line_sys_mpc_continuous}
\begin{eqnarray}
\dot{x} &\approx& A_c(x-\hat{x}_{k}) + B_c(u-u_{k-1})+e_c\label{line_sys_mpc_continuous_x}\\
y &\approx& C(x-\hat{x}_{k})+h_c
\end{eqnarray}
\end{subequations}
where $A_c=\frac{\partial f}{\partial x}|_{\hat{x}_k, u_{k-1}, \hat{d}_k}$, $B_c=\frac{\partial f}{\partial u}|_{\hat{x}_k, u_{k-1}, \hat{d}_k}$, $e_c=f(\hat{x}_k,u_{k-1},\hat{d}_k)$, $C=\frac{\partial g}{\partial x}|_{\hat{x}_k, \hat{d}_k}$ and $h_c=g(\hat{x}_k,\hat{d}_k)$. Here the differential eqn (\ref{line_sys_mpc_continuous_x}) has to be discretised for obtaining a discrete-time model. The discretization method includes the exact and the approximated  discretization. One common discretization method is the the one-step Euler's approximate method, then a discrete-time model is obtained as follows
\begin{subequations}\label{line_sys_mpc}
\begin{eqnarray}
x_{k+1} &=& A x_{k} +B u_k + e\label{line_sys_mpc_x} \\
y_k &=& C x_k + h\label{line_sys_mpc_y}
\end{eqnarray}
\end{subequations}
where $e=t_s\left(e_c-A_c \hat{x}_k-B_c u_{k-1}\right)$, $A=I+t_s A_c$, $B=t_s B_c$, $h=h_c-C \hat{x}_k$. Then, a MPC tracking problem formulation is listed as follows
\begin{eqnarray}\label{ss_mpc}
\min && \frac{1}{2}\sum_{k=0}^{T-1}\left\|\left(y_{k+1}-r_{k+1}\right)\right\|_{W^{y}}^{2}+\left\| \Delta u_{k}\right\|_{W^{\Delta u}}^{2}\nonumber\\
\text {s.t.} &&\text{Eqn}\quad(\ref{line_sys_mpc_x}), k=0,\ldots,T-1\nonumber\\
&& \text{Eqn}\quad(\ref{line_sys_mpc_y}), k=1,\ldots,T\nonumber\\
&& u_k =  u_{k-1} + \Delta u_{k}, k=0,\ldots,T-1\nonumber\\
&& y_{\min } \leq y_{k} \leq y_{\max }, k=1,\ldots,T\nonumber\\
&& u_{\min } \leq u_{k} \leq u_{\max }, k=0,\ldots,T-1\nonumber\\
&& \Delta u_{\min } \leq \Delta u_{k} \leq \Delta u_{\max }, k=0,\ldots,T-1\nonumber\\
&& x_0 = \hat{x}_{k}
\end{eqnarray}
The MPC tracking problem~(\ref{ss_mpc}) can be solved by many quadratic programming (QP) algorithms, such as the active-set, the interior-point and the ADMM-based solver. Note that the model parameters $\{A,B,e,C,h\}$ in  (\ref{line_sys_mpc}) are related to the recursively updated estimates $\{\hat{x}_k,\hat{d}_k\}$ and the last control input $u_{k-1}$, it means that the MPC tracking problem~(\ref{ss_mpc}) has to be reformulated at each sampling time. Most QP solvers requires an explicit condensing or sparse MPC-to-QP construction. In such a SL-MPC scenario, the MPC-to-QP construction has to be performed online at each sampling time, as does solving the QP problem. Indeed, often the online MPC-to-QP construction has a comparable computational cost to solving the QP problem itself online, especially when warm-starting strategies are employed. Our previous work \cite{wu2021simple}, a construction-free \textit{CDAL} solver, can avoid explicit MPC-to-QP construction in state-space MPC problem.

\section{Interpretative and Adaptive MPC}
Compared to the exact NMPC method, the approximate method, SL-MPC with EKF, could significantly reduce the online computational burden for nonlinear output-feedback MPC problems. Nonetheless, the SL-MPC with EKF involves two-times linearizations, and its embedded implementation, including integrator, linearizer, EKF and time-changing MPC problem solver, remains a difficult task for control engineers. 

In the SL-MPC with EKF method, the model parameters $\{A,B,e,C,h\}$ of the model (\ref{ss_mpc}) are dependant to the last control input $u_{k-1}$ and the $\{\hat{x}_k,\hat{d}_k\}$, which are recursively estimated by the EKF algorithm. In a sense, the model are recursively updated (or feedback corrected) by the EKF algorithm. And our previous work \cite{wu2022equivalence} illustrates that an observable linear state-space (SS) model can be equivalently transformed into an ARX model, which will be also introduced in the following subsection \ref{sec:equivalent}. Adding an SS-to-ARX transformation step in SL-MPC with EKF method would not affect the closed-loop control performance. However, an additional SS-to-ARX transformation step relies on the choosing observer design and additional matrix-matrix multiplication cost, which makes it no advantage in the online SL-MPC with EKF method. This inspired us to update the ARX model directly by the EKF algorithm from the streaming input-output data, to avoid an explicit SS-to-ARX transformation. By viewing the ARX model parameters as states of a system, the EKF algorithm could recursively updates the ARX model parameters. In fact, a system divides the states and parameters according to their changibng rate, states change fast and parameters change slowly. And this ARX model identification and tracking by the Kalman Filter based algorithm has been widely used in adaptive control \cite{guo1990estimating}. 

\begin{figure}
\centering
        \hspace*{-1em}\includegraphics[width=0.8\columnwidth]{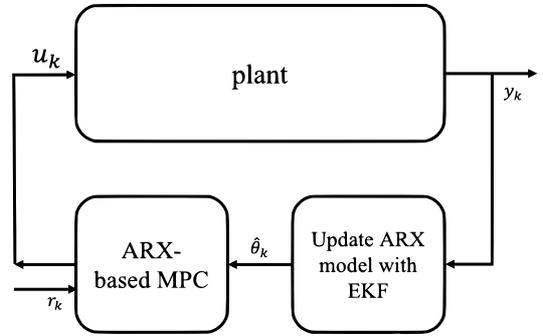} 
        \caption{Schematic diagram of interpretative and adaptive MPC}
        \label{IA_MPC} 
\end{figure}

As the EKF algorithm requires a sufficiently small error in initial estimates, and the data-driven identification (initialization) of ARX model is black-box without interpretability like first-priciple-based model, we propose the introduction of the off-line SS-to-ARX transformation in our interretative and adaptive MPC \textit{IA-MPC} framework. The main steps of our \textit{IA-MPC} framework are list in Tab. \ref{tab:ia-mpc}. Its \textit{Step} 1 to 3 are performed offline, and the \textit{Step} 1 is to obtain a linear state-space model by linearizing a first-principle-based model at initial point; the optional \textit{Step} 2 is to obtain a minimal state-space realization by using model reduction when the state dimension is large such as in distributed parameter models or computational fluid dynamic models; After possible state elimination, the \textit{Step} 3 is to obtain an equivalent ARX model by designing an SS-to-ARX transformation. In addition to the gained interpretability and consistent initial estimates, this ARX model acquisition also avoids the difficulty in choosing the ARX model orders in data-driven ARX model identification. The detailed equivalent SS-to-ARX transformation will be illustrated in the following subsection \ref{sec:equivalent}. The \textit{Step} 4 is the online closed-loop MPC control, in which the ARX model parameters are updated by the EKF algorithm from streaming input-output data (see the subsection \ref{sec:arx_ekf}) and an corresponding ARX-based MPC problem are solved by our previous construction-free \textit{CDAL-ARX} algroithm \cite{wu2022construction} (see the subsection \ref{sec:arx_mpc}). For a comparison with the traditional SL-MPC with EKF framework shown in Fig. \ref{EKF_MPC}, the schematic diagram of our \textit{IA-MPC} framework is showed in Fig. \ref{IA_MPC}.
\begin{table}[!htbp]
    \caption{Interpretative and Adaptive MPC framwork}
    \centering
    \begin{tabular}{ll}
    \hline\hline
        Step &  Detailed description\\
    \midrule
    1 & \begin{tabular}{l}
        obtain a linear state-space model based on \\
        the linearization of a first-principle-based\\
        model at initial point
        \end{tabular}\\
    \midrule
    2(optional) & \begin{tabular}{l}
        obtain a minimal realization by using model \\
        reduction when states dimension is large such as\\ in distributed
        parameter models or\\
        computational fluid dynamic models
        \end{tabular}\\
    \midrule
    3   & \begin{tabular}{l}
        design an SS-to-ARX transformation to obtain\\
        an equivalent ARX model (robust to process \\
        and measurement noises)
        \end{tabular}\\
    \midrule
    4   & \begin{tabular}{l}
        combine the online EKF update of ARX models\\
        and ARX-based MPC algorithms to be used in\\
        closed-loop simulations
        \end{tabular}\\
    \hline\hline  
    \end{tabular}\label{tab:ia-mpc}
\end{table}

\subsection{Equivalent SS-to-ARX transformation}\label{sec:equivalent}
In our preivous work \cite{wu2022equivalence}, we conclude three SS-to-ARX transformations, namely the Cayley-Hamilton-based, Observer-Theory-based and Kalman-Filter-based transformations. The Cayley-Hamilton-based tranformation can derive the unique and equivalent ARX model but with sensitivity to noises. The Observer-Theory-based and Kalman-Filter-based transformations has the same structure in the transformed ARX model, and the Kalman-Filter-based transformation requires a larger order than the Observer-Theory-based transformation. In fact, the order determines the noises robustness of the ARX model, which can easily tuned by the pole placement method. Clearly, larger order brings better noises robustness but also leads to more computational cost in ARX model update and ARX-based MPC. Herein we mainly present how to transform the adaptive linearized state-space model (\ref{line_sys_mpc}) with the help of the the Observer-Theory-based transformation.

By utilizing a gain matrix $L$ in the SS model (\ref{line_sys_mpc}) to begin the derivation of the SS-to-ARX transformation as follows
\begin{equation}\label{SS_xyu}
\begin{aligned}
    x_{k+1} &= A x_k + B u_k  + e - L y_k + L y_k\\
    &=(A-LC)x_t+B u_t + L y_t + e
\end{aligned}
\end{equation}
Based on the above evolution equation (\ref{SS_xyu}), calculating from previous time $t-p$ to current time $t$ can lead to the following equation
\begin{equation}\label{SS_p_t}
\begin{aligned}
    x_{k}&=(A-LC)^p x_{k-p} +\sum_{i=1}^{p}(A-LC)^{i-1} L y_{k-i} \\ 
    &+ \sum_{i=1}^{p}(A-LC)^{i-1} B u_{k-i} + \sum_{i=1}^{p}(A-LC)^{i-1}e
\end{aligned}
\end{equation}
If a gain matrix $L$ exists such that $(A-LC)^p$ vanishes to zero, i.e.,
\begin{equation}\label{condition_ALC}
(A - LC)^{k} \equiv 0, \quad k \geq p
\end{equation}
From linear system theory, the existence of such an gain matrix $L$ is assured as long as the system is observable. Thus, the term $(A-LC)^p x_{t-p}$ in (\ref{SS_p_t}) is zero for $k\geq p$, then multiplying the matrix $C$ on both sides of the equation (\ref{SS_p_t}) can derive the following ARX model,
\begin{equation}\label{OT-based SS-to-ARX}
    y_k=\sum_{i=1}^{p}\Psi_i y_{k-i}+\sum_{i=1}^{p}\Omega_i u_{k-i}+\zeta
\end{equation}
where $\Psi_i=C(A-LC)^{i-1} L$, $\Omega_i=C(A-LC)^{i-1} B$ and $\zeta=\sum_{i=1}^{k}C(A-LC)^{i-1} e$

Next, we provide the analysis to tell why the gain matrix $L$ can be viewed as an observer gain. The original state-space model~(\ref{line_sys_mpc}) has an observer gain $L$ of the following form
\begin{equation}\label{OT_SS_to_ARX}
\begin{aligned}
\hat{x}_{k+1} &= A\hat{x}_k + Bu_k + L(y_k-\hat{y}_k)+e\\
\hat{y}_k &= C\hat{x}_k
\end{aligned}
\end{equation}
where $\hat{x}_k$ is the estimated state. The state estimation error can be denoted as $\epsilon_k=x_k-\hat{x}_k$, then $\epsilon_k$ follows the dynamic equation,
\begin{equation}
    \epsilon_{k+1} = (A-LC)\epsilon_k
\end{equation}
The estimated state $\hat{x}_k$ will converge the actual value $x_k$ as $k$ tends to infinity if the matrix $A-LC$ is asymptotically stable, namely the condition (\ref{condition_ALC}), which can be satisfied by using the pole placement method.

\subsection{ARX model update with decoupled EKF algorithm}\label{sec:arx_ekf}
By viewing the time-varying ARX model parameters as system states, the ARX model (\ref{OT-based SS-to-ARX}) is re-formulateed as the following state-space form
\begin{subequations}
\begin{eqnarray}
&\theta_{k}=\theta_{k-1}+w_{k}\label{eqn:random_walk_model}\\
&y_{k}=\varphi_{k}^{\prime}\theta_{k}+v_{k}\label{OT_based SS-to-ARX y}
\end{eqnarray}
\end{subequations}
where $\{w_{k}\}$ is a sequence of independent random vectors and $\{v_{k}\}$ is a output noise sequence. Although the variances of $\{w_{k}\}$ and $\{v_{k}\}$ are unknown in real applications and the actual parameters $\{\theta_{k}\}$ may change differently from the above random walk model (\ref{eqn:random_walk_model}), the EKF algorithm can still work very well \cite{cao2004analysis}.

Let's write the Eqn (\ref{OT_based SS-to-ARX y}) as the following separated formulation
\begin{equation}
    y_k(j) = \varphi_k^{\prime} \theta_k(j) + v_k(j),j=1,\ldots,n_y
\end{equation}
where $n_y$ denotes the dimensions of $y$, $y_k(j)$ denotes the $j$-th element of $y_k$. $\theta_k$ and $\varphi_k$ have the following relationship according to (\ref{OT-based SS-to-ARX})
\begin{equation}
    \begin{aligned}
    &\theta_k=\left[\begin{array}{cccc}
    \theta_k(1)^{\prime} & \theta_k(2)^{\prime} &
    \ldots &
    \theta_k(n_y)^{\prime}
    \end{array}\right]^{\prime}\\
    &\theta_k(j)^{\prime}=\small\left[\Psi_1(j,:), \ldots, \Psi_p(j,:), \Omega_1(j,:), \ldots, \Omega_p(j,:), \zeta(j)\right]\\
    &\varphi_k^{\prime}=\small\left[y_{k-1}^{\prime}, \ldots, y_{k-p}^{\prime}, u_{k-1}^{\prime}, \ldots, u_{k-p}^{\prime}, 1\right]
    \end{aligned}
\end{equation}
The advantage of the above separated ARX formulation is that it allows the EKF algorithm to run in a parallel way and avoids the matrix-inverse operation, only involving scalar division, compared to the EKF in the SL-MPC (see Tab. \ref{tab:EKF-NMPC}). We list the EKF computation scheme of our \textit{IA-MPC} framework in Tab. \ref{tab:ekf_updates_ARX}.

\begin{table}[!htbp]
    \caption{EKF for ARX model updating}
    \centering
    \begin{tabular}{ll}
    \hline\hline% \toprule
        Step &  Formula\\
    \midrule
        Initialization & $P_0=\epsilon I$, $\epsilon$ is a large number,\\
        & initial value $\theta(1),\ldots,\theta(n_y)$ from Step 3 in Tab. \ref{tab:ia-mpc},\\
        & $Q$ is the process noise covariance,\\
        & $r$ is the measurement noise covariance.\\
        \midrule
        1-Prediction &
        $\hat{\theta}_k^{-}(j)=\hat{\theta}_{k-1}(j),j=1,\ldots,n_y$\\
        \\
        2-Correction &
        $\hat{\theta}_k(j)=\hat{\theta}_{k}^{-}(j)+K_k\left(y_k(j)-\varphi_k^{\prime}\hat{\theta}_{k}^{-}(j)\right),$\\
        & $j=1,\ldots,n_y$\\
        \midrule
        \multicolumn{2}{c}{\begin{tabular}{l}
        $P_k^{-}=P_{k-1}+Q$\\
        \quad
        \end{tabular}}\\
        \multicolumn{2}{c}{$K_k=\frac{1}{\varphi_k^{\prime}P_k^{-}\varphi_k+r}
        P_{k}^{-}\varphi_k$}\\
        \quad \\
        \multicolumn{2}{c}{$P_k=\left(I-K_k\varphi_k^{\prime}\right)P_k^{-}$}\\
    \hline\hline% \bottomrule    
    \end{tabular}
    \label{tab:ekf_updates_ARX}
\end{table}

% And, the EKF also provides the advantage of individually determined time variation of states. 

\subsection{Construction-free ARX-based MPC algorithm}\label{sec:arx_mpc}
In our \textit{IA-MPC} framework, the ARX model is updated by the EKF algorithm at each sampling time. Thus it leads to the corresponding time-changing ARX-based MPC tracking problem as follows.,
\begin{eqnarray}\label{ARX_mpc}
\min && \frac{1}{2}\sum_{k=0}^{T-1}\left\|\left(y_{k+1}-r_{k+1}\right)\right\|_{W^{y}}^{2}+\left\| \Delta u_{k}\right\|_{W^{\Delta u}}^{2}\nonumber\\
\text {s.t.} && y_k=\sum_{i=1}^{p}\hat{\Psi}_i y_{k-i}+\sum_{i=1}^{p}\hat{\Omega}_i u_{k-i}+\hat{\zeta}, k=1,\ldots,T\nonumber\\
&& u_k =  u_{k-1} + \Delta u_{k}, k=0,\ldots,T-1\nonumber\\
&& y_{\min } \leq y_{k} \leq y_{\max }, k=1,\ldots,T\nonumber\\
&& u_{\min } \leq u_{k} \leq u_{\max }, k=0,\ldots,T-1\nonumber\\
&& \Delta u_{\min } \leq \Delta u_{k} \leq \Delta u_{\max }, k=0,\ldots,T-1
\end{eqnarray}
In such a a time-changing setting, the computation time spent in constructing its optimization problem and solving itself should be together considered. Indeed, often the online MPC-to-QP construction has a comparable computational cost to solving the MPC problem itself online, especially when warm-starting strategies are employed. Our previous construction-free \textit{CDAL-ARX} algorithm \cite{wu2022construction} can avoid the explicit MPC-to-QP construction, thus allowing it to be suitable for this scenario. 

In each iteration of our \textit{CDAL-ARX} algorithm, an augmented Lagrangian (AL) subproblem is solved by the coordinate descent (CD) method, and the accelerated Nesterov’s scheme is used to update the Lagrangian dual variables. In particular, an efficient coupling scheme between CD and AL method is proposed to exploit the structure of the ARX-MPC problem, which can reduce the computation cost of each inner iteration. The detailed description of algorithm implementation can be found in \cite{wu2022construction}. In addition to the notable construction-free feature, our \textit{CDAL-ARX} is also matrix-free and library-free which makes it practically useful in embedded deployment. 

\section{Numerical Examples}
In this section, we test the performance of our proposed IA-MPC method against other two methods, namely the nonlinear MPC method with EKF and the SL-MPC with EKF method. Their MPC solution are based on CasADi v3.5.5\cite{Andersson2019} and our previous CDAL algorithm for state-space model based MPC \cite{wu2021simple}, respectively, except for the same EKF computational procedure. The reported simulations are executed in MATLAB R2020a on a MacBook Pro with 2.7~GHz 4-core Intel Core i7 and 16GB RAM. Four typical nonlinear MPC numerical examples are used to investigate whether our IA-MPC method works well and provide comparisons with traditional methods. 

\subsection{Problem descriptions}
\begin{enumerate}
\item \textit{Two tank }problem: we consider the cascaded two tanks system, which is a fluid level control system. The input signal controls the water pump that pumps the water from a reservoir into the upper water tank. The water of the upper water tank also flows through a small opening into the lower water tank. The water of the lower water tank flows through a small opening into the reservoir. Herein without considering the overflow effect, we adopt the Bernoulli’s principle and conservation of mass to derive the following first-principle model:
\begin{equation}\label{eqn:two_tank}
\begin{aligned}
\dot{x}_{1} &=-k_{1} \sqrt{x_{1}}+k_{2} u \\
\dot{x}_{2} &=k_{1} \sqrt{x_{1}}-k_{3} \sqrt{x_{2}} \\
y &=x_{2}
\end{aligned}
\end{equation}
where $x_1$ and $x_2$  are the water level of the upper and lower water tank, respectively. The full states cannot be measured only the $x_2$ as the measured output $y$. $u$ is the input signal, and $k_1$ , $k_2$ and $k_3$ are constants depending on the system properties, herein we adopt value $0.5$ for all of them. The sampling time of discrete digital control is $0.2$ s and the control goal is to make the output $y$ track the given reference signal subject to the input constraints $0 \leq u \leq 2$ and the input increment constraints $-0.5 \leq \Delta u \leq 0.5$.
\item \textit{Bilinear motor} problem: one common nonlinear control benchmark is a bilinear DC motor plant, whose equation is described as follows,
\begin{equation}\label{eqn:bilinear_motor}
\begin{aligned}
\dot{x}_{1} &=-\left(R_{a} / L_{a}\right) x_{1}-\left(k_{m} / L_{a}\right) x_{2} u+u_{a} / L_{a} \\
\dot{x}_{2} &=-(B / J) x_{2}+\left(k_{m} / J\right) x_{1} u-\tau_{l} / J \\
y &=x_{2}
\end{aligned}
\end{equation}
where $x_1$ and $x_2$ are the rotor current and angular velocity, respectively. The full states cannot be measured only the $x_2$ as the measured output $y$. The control input $u$ is the stator current. The system parameters are $L_a = 0.314$, $R_a = 12.345$, $k_m = 0.253$, $J = 0.00441$, $B = 0.00732$, $\tau_{l}= 1.47$, $u_a= 60$. The bilinearity of (\ref{eqn:bilinear_motor}) appears between the state and the control input. The sampling time of discrete digital control is $0.01$ s and the control goal is to make the output $y$ track the given reference signal subject to the input constraints $0 \leq u \leq 2$ and the input increment constraints $-1 \leq \Delta u \leq 1$. 
\item \textit{CSTR} problem: one typical nonlinear process control benchmark is a continuous stirred tank reactor (CSTR) problem, which is described by the following continuous-time nonlinear model,
\begin{equation}\label{eqn:CSTR}
\begin{aligned}
\dot{C_A} &= C_{A,i}-C_A-k_0 e^{\frac{-EaR}{T}}C_A \\
\dot{T} &= T_{i} + 0.3 T_c - 1.3T + 11.92 k_0 e^{\frac{-EaR}{T}}C_A\\
y &= T
\end{aligned}
\end{equation}
where $C_A$ is the concentration of reagent A, $T$ is the temperature of the reactor, $C_{A,i}$ is the inlet feed stream concentration, which is assumed to have the constant value $10.0$ kgmol/m$^3$. The unmeasured disturbance comes from the inlet feed stream temperature $T_{i}$, which has slow fluctuations represented by $T_{i} = 298.15 + 5 \sin(0.05 t)$ $K$. The manipulated variable is the coolant temperature $T_c$. The constants $k_0 = 34930800$ and $EaR = -5963.6$ (in MKS units). The sampling time of discrete digital control is $0.5$ s and the control goal is to manipulates the coolant temperature $T_c$ to track a higher temperature of the reactor (equals a higher conversion rate) as well as reject the unmeasured disturbance $T_{i}$. The physical constraints comes from the input increment constraints $-1 \leq \Delta T_c \leq 1$. Note that the unmeasured disturbance $T_i$ is estimated by the EKF algorithm in SL-MPC and NMPC methods.
\item \textit{Van der Pol oscillator} problem: we consider the nonlinear Van der Pol oscillator with a time varying parameter, and its dynamics are given by,
\begin{equation}\label{eqn:VanderPol}
\begin{aligned}
\dot{x} &= v \\
\dot{v} &= \mu(t) (1-x^2)v-x+u \\
\end{aligned}
\end{equation}
where $x$ is the position, $v$ is the velocity, $u$ is the control input and $\mu(t)$ is a piecewise function defined as
\begin{equation*}
\mu(t)= \begin{cases}1 & \text { if } t \leq 50 \\ 3 & \text { if } t>50\end{cases}
\end{equation*}
The sampling time of discrete digital control is $0.2$ s and the control goal is to make the output $y$ track the given reference signal subject to the input constraints $-10 \leq u \leq 10$ and the input increment constraints $-10 \leq \Delta u \leq 10$. Note that the unmeasured time-varying parameter $\mu(t)$ is estimated by the EKF algorithm in SL-MPC and NMPC methods.
\end{enumerate}

\subsection{Problems settings}
\begin{figure*}[htbp]
\centering 
% \flushleft
\subfigure[Tracking performance of IA-MPC method without noise]{
\begin{minipage}{8cm}\label{fig_twotank_a}
\centering
\includegraphics[scale=0.45]{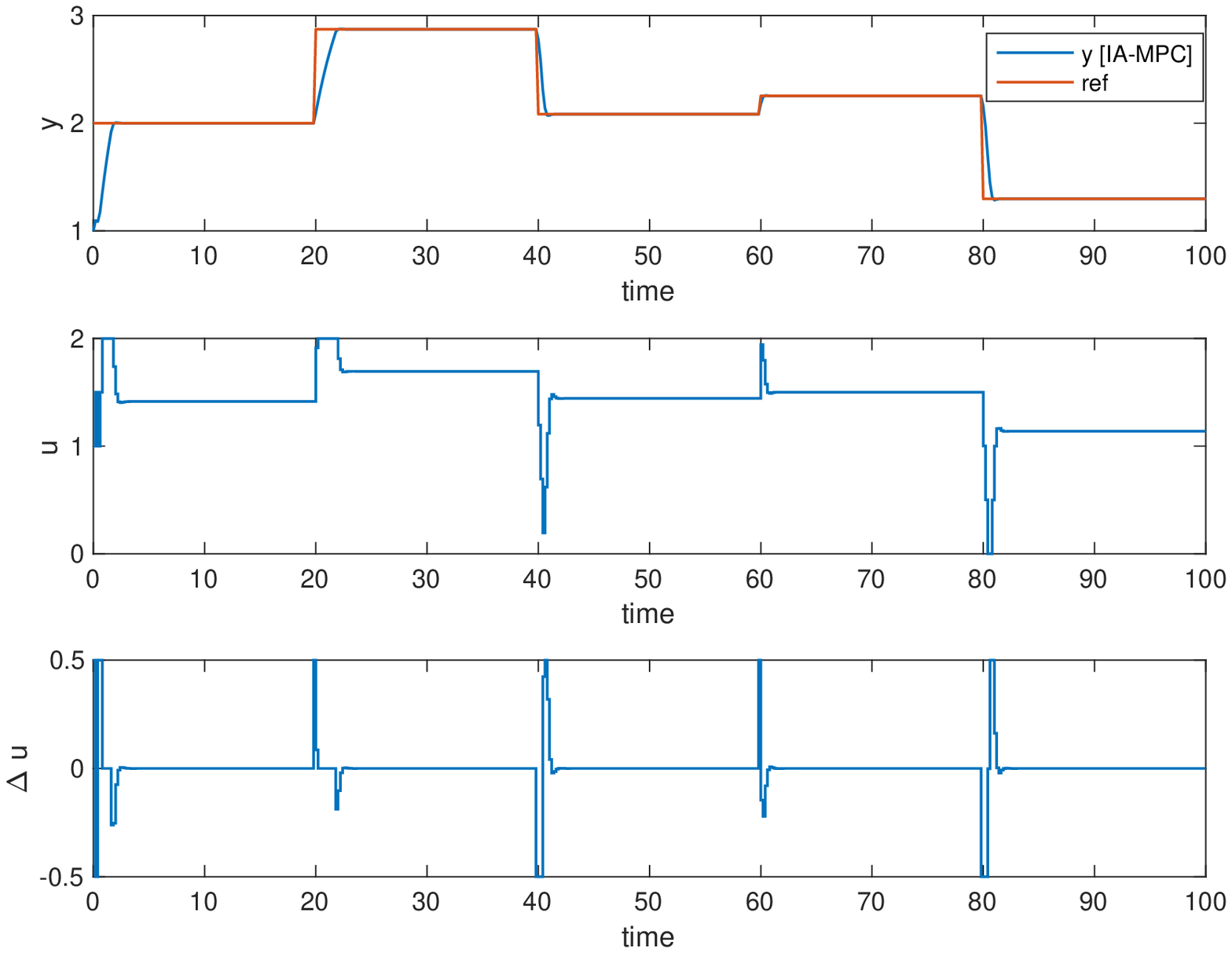}
\end{minipage}
}\subfigure[Comparison of IA-MPC, SL-MPC, and NMPC methods with noise]{
\begin{minipage}{8cm}\label{fig_twotank_b}
\centering
% \flushleft
\includegraphics[scale=0.45]{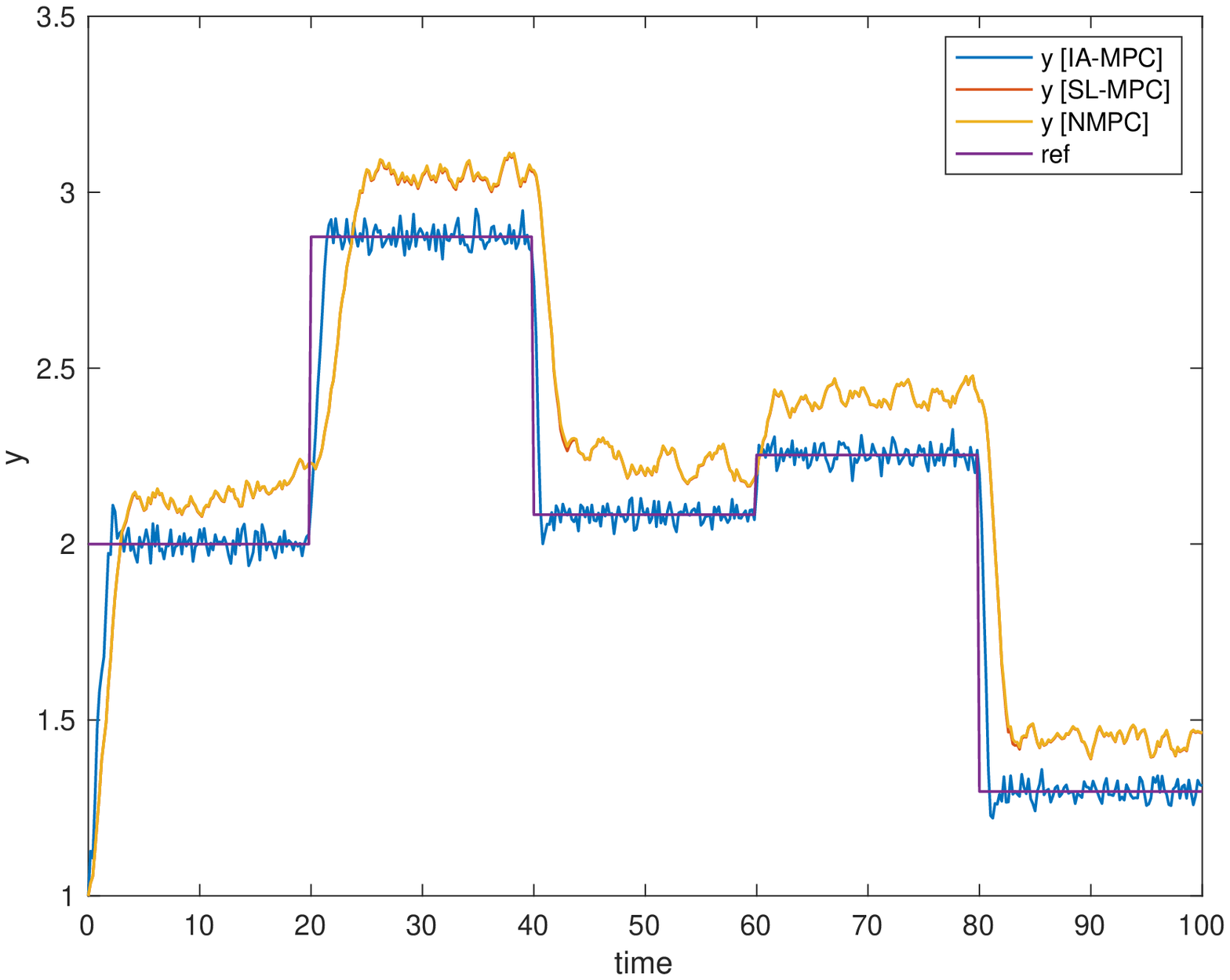}
\end{minipage}
}
\caption{The closed-loop control performances in the \textit{Two tank} problem}
\label{fig_twotank}
\end{figure*}

\begin{figure*}[htbp]
\centering 
% \flushleft
\subfigure[Tracking performance of IA-MPC method without noise]{
\begin{minipage}{8cm}\label{fig_bilinear_a}
\centering
\includegraphics[scale=0.45]{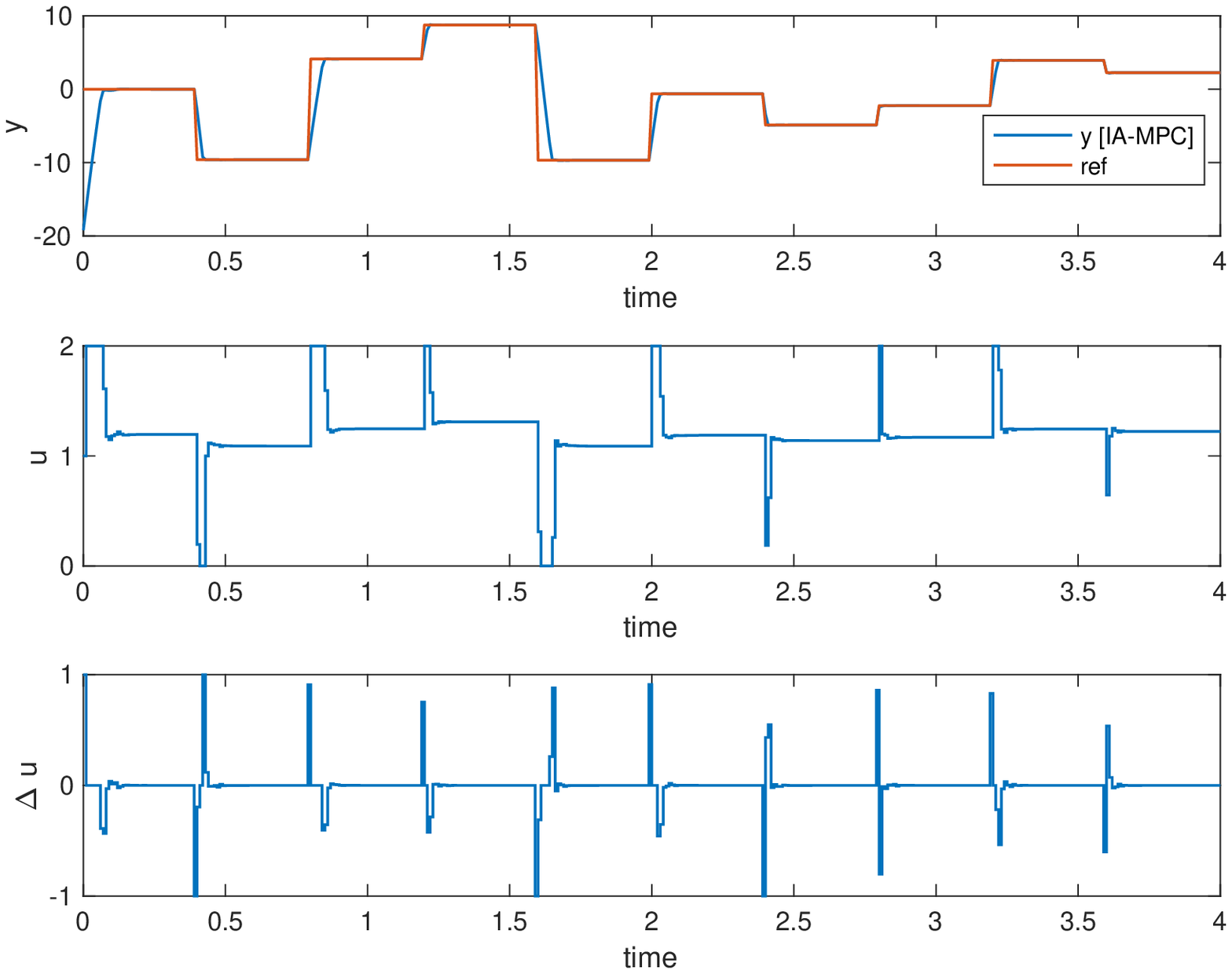}
\end{minipage}
}\subfigure[Comparison of IA-MPC, SL-MPC, and NMPC methods with noise ]{
\begin{minipage}{8cm}\label{fig_bilinear_b}
\centering
% \flushleft
\includegraphics[scale=0.45]{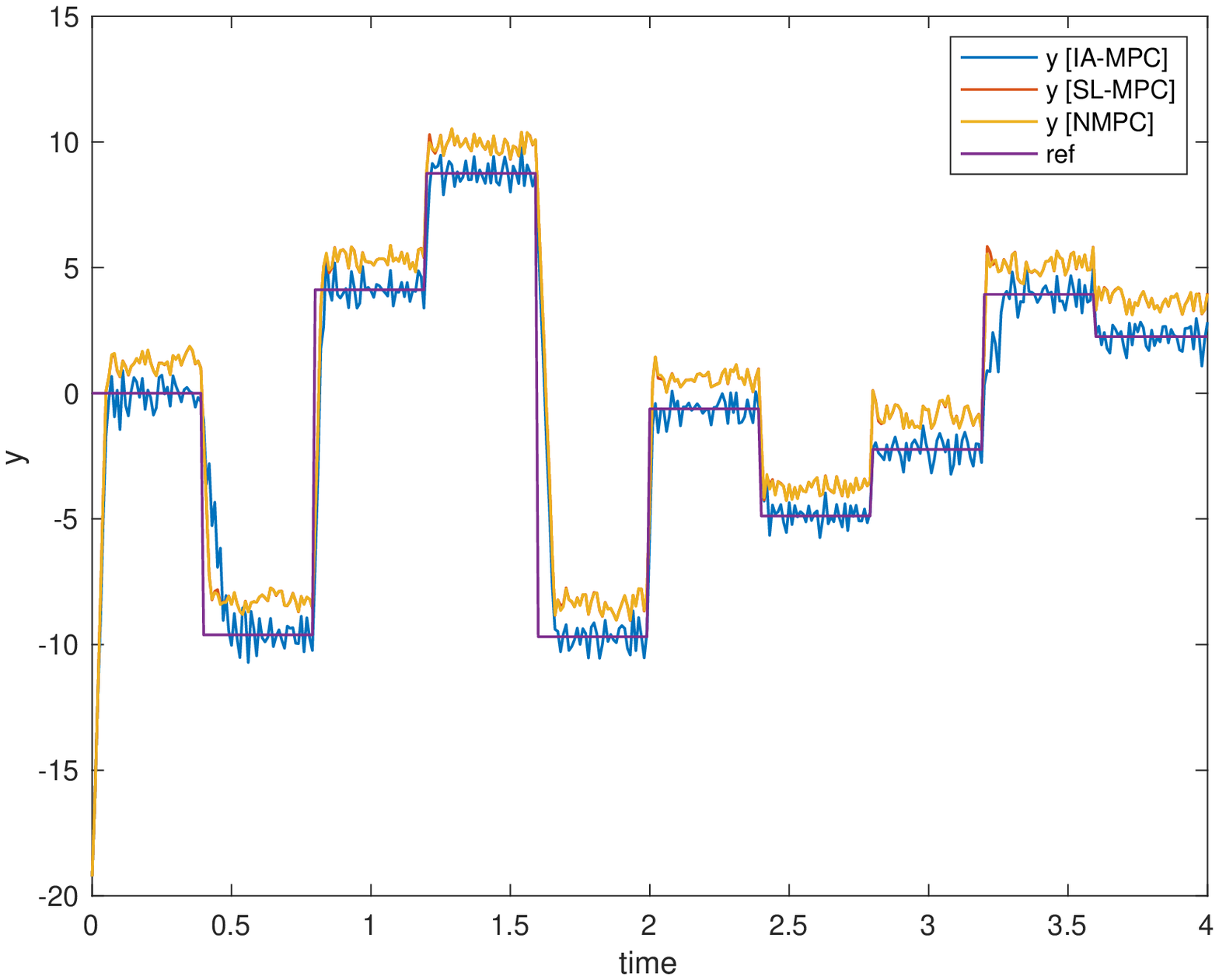}
\end{minipage}
}
\caption{The closed-loop control performances in the \textit{Bilinear motor} problem}
\label{fig_bilinear}
\end{figure*}

\begin{figure*}[htbp]
\centering 
% \flushleft
\subfigure[Comparison without noise]{
\begin{minipage}{8cm}\label{fig_CSTR_a}
\centering
\includegraphics[scale=0.45]{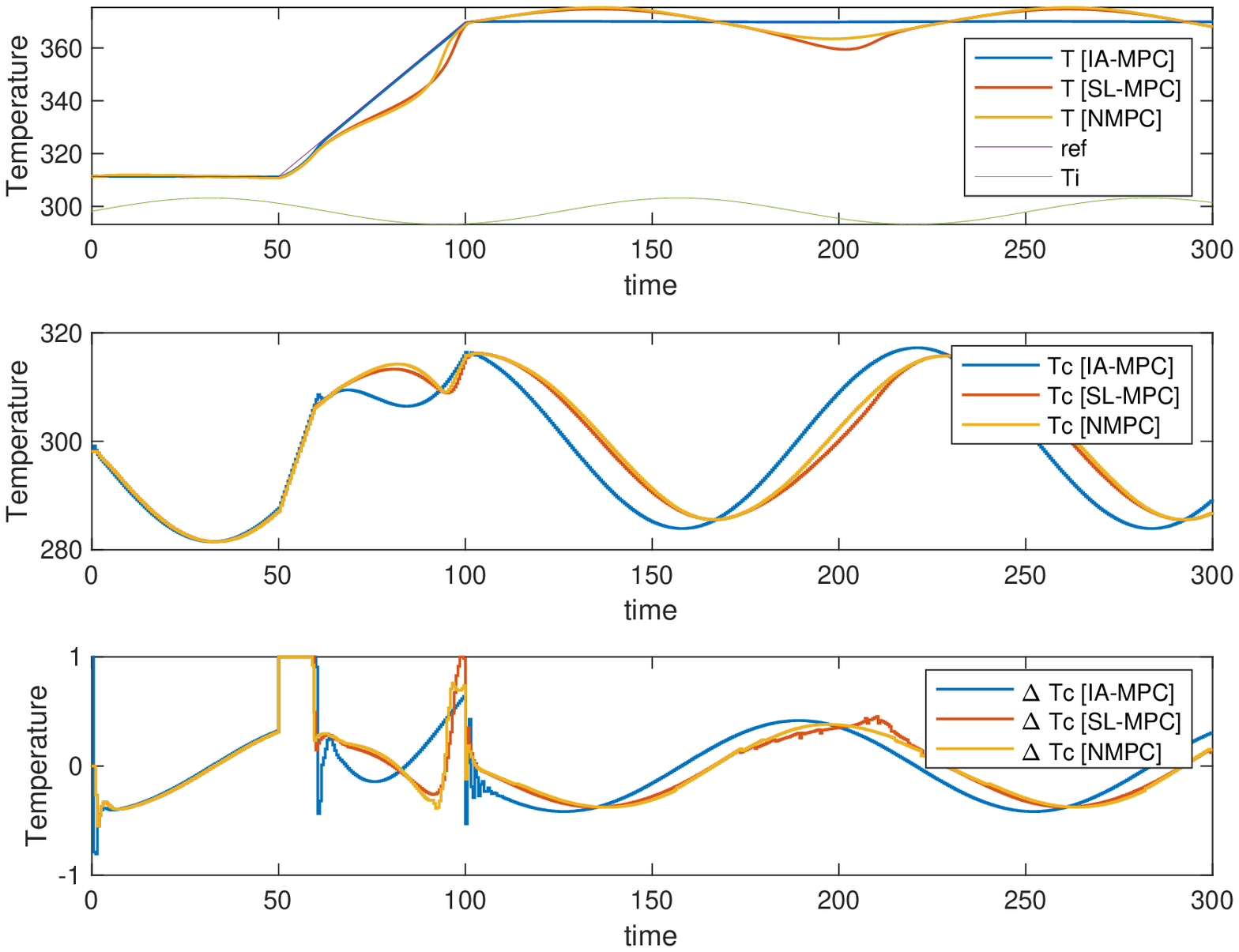}
\end{minipage}
}\subfigure[Comparison with noise ]{
\begin{minipage}{8cm}\label{fig_CSTR_b}
\centering
% \flushleft
\includegraphics[scale=0.45]{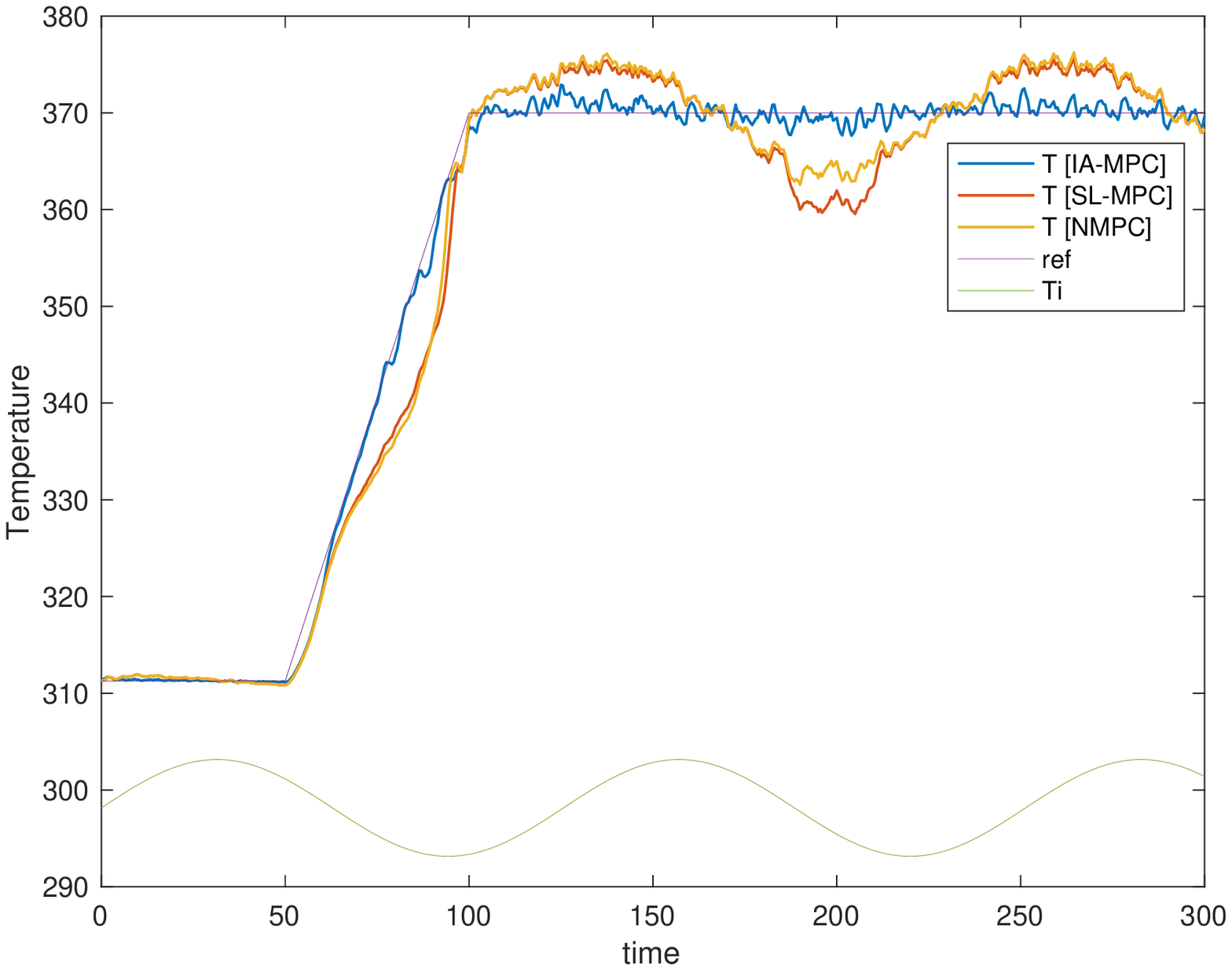}
\end{minipage}
}
\caption{The closed-loop control performances in the \textit{CSTR} problem}
\label{fig_CSTR}
\end{figure*}

\begin{figure*}[htbp]
\centering 
% \flushleft
\subfigure[Comparison without noise]{
\begin{minipage}{8cm}\label{fig_VanderPol_a}
\centering
\includegraphics[scale=0.45]{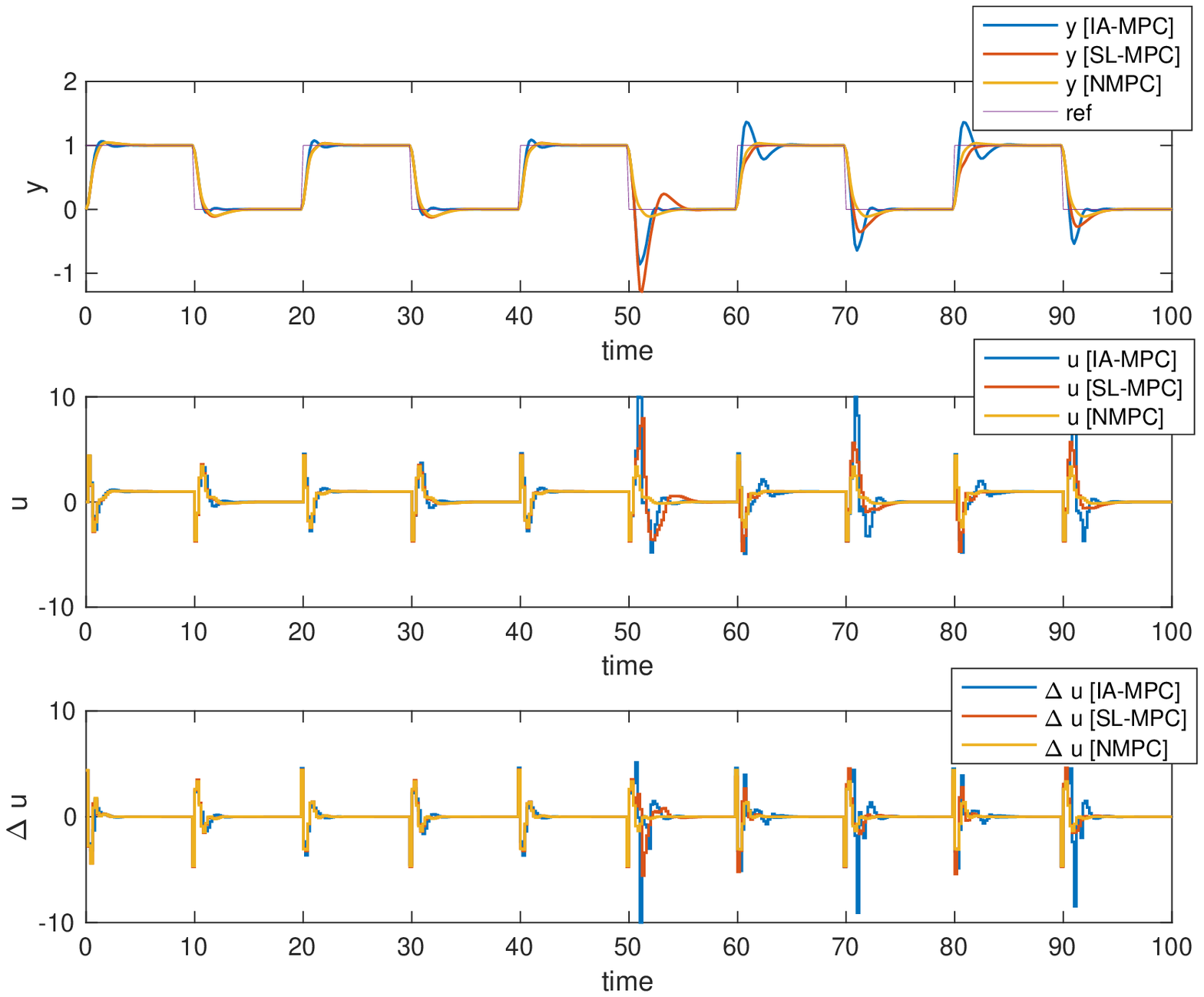}
\end{minipage}
}\subfigure[Comparison with noise]{
\begin{minipage}{8cm}\label{fig_VanderPol_b}
\centering
% \flushleft
\includegraphics[scale=0.45]{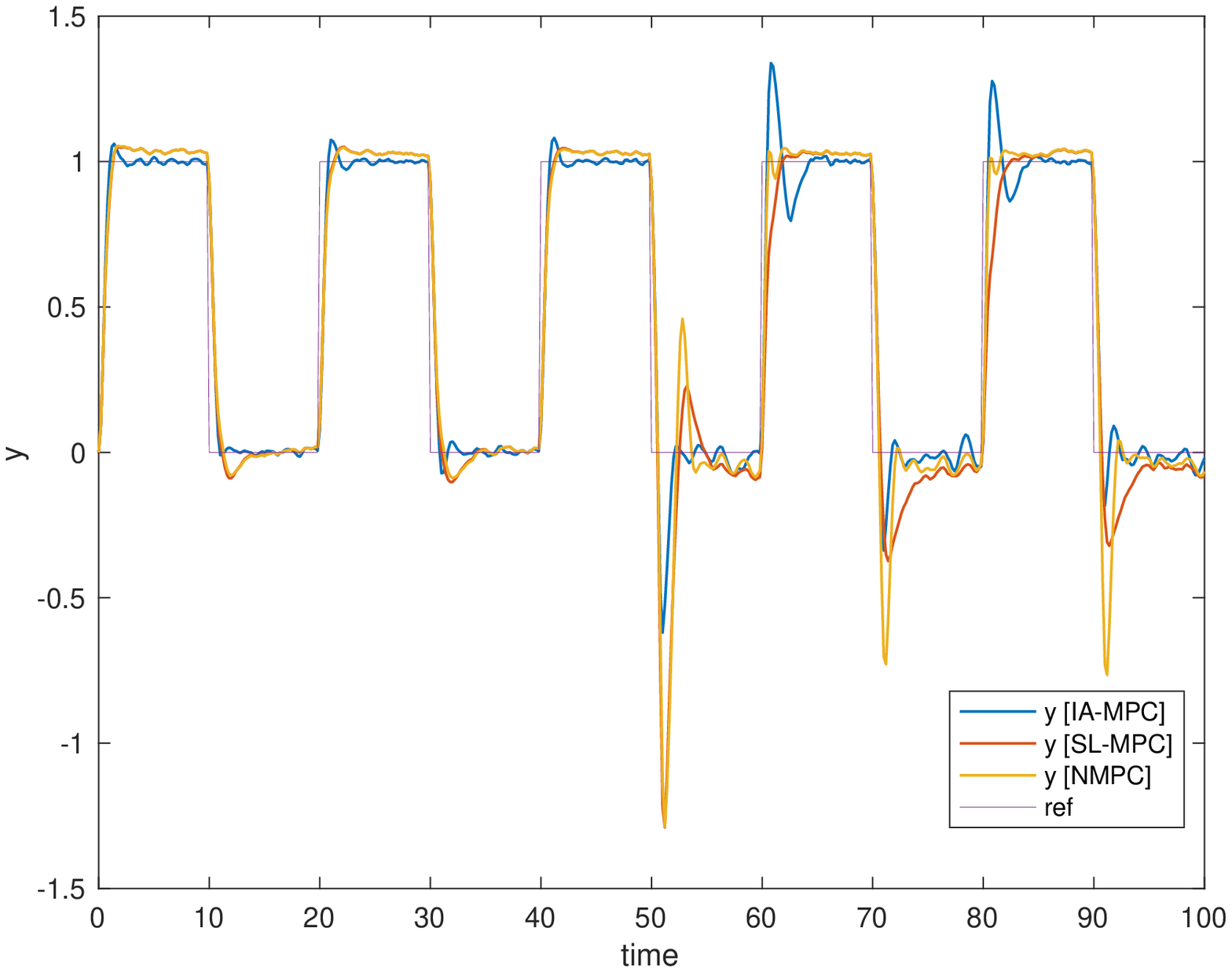}
\end{minipage}
}
\caption{The closed-loop control performance in the \textit{Van der Pol} problem}
\label{fig_VanderPol}
\end{figure*}

The above four problems share the same MPC and EKF settings: the MPC prediction horizon $N_p=10$, the MPC output weight $W_y=10$ and the MPC input increment weight $W_{\Delta u}=0.1$, the initial error covariance matrix of EKF $P_0=10 I$, the process noise covariance matrix of EKF $Q=0.01 I$ and the measurement noise covariance of EKF $R=0.01$. We consider two simulation scenarios, that is without and with process noises, respectively.
\begin{enumerate}
    \item \textit{Two tank} problem: its process noise scenario is given a random noise $0.05 \times$rand(2,1), and the initial conditions are $x_1 = 1$, $x_2 = 1$ and $u = 1$ for obtaining the linearized state-space model, which is then transformed into the ARX model with order $p=3$ by using poles $[0.01, 0.02]$. Its tracking signal is randomly selected every $20$ s in the range $[1,3]$.
    \item \textit{Bilinear motor} problem: its process noise scenario is given a random noise $1 \times$rand(2,1), and the initial conditions are $x_1 = 5.2542$, $x_2 = -19.2205$ and $u = 1$ for obtaining the linearized state-space model, which is then transformed into the ARX model with order $p=5$ by using poles $[0.05, 0.1]$. Its tracking signal is randomly selected every $0.4$ s in the range $[-10,10]$. 
    \item \textit{CSTR} problem: its process noise scenario is given a random noise $0.1 \times$rand(2,1), and the initial state of the reactor is at a low conversion rate, with $C_A = 8.57$ kgmol/m$^3$, $T = 311$ K, and then the linearized state-space model around the initial conditions is transformed into the ARX model with order $p=3$ by using poles $[0.01, 0.02]$. Its tracking signal gradually changes from $311.2639$ K to $370$ K during $50$~$100$ s and then holds constant.
    \item \textit{Van der Pol oscillator} problem: its process noise scenario is given a random noise $1 \times$rand(2,1), and the initial conditions are $x_1 = 0$, $x_2 = 0$ and $u = 0$ for obtaining the linearized state-space model, which is then transformed into the ARX model with order $p=3$ by using poles $[0.005, 0.01]$. Its tracking signal switches between $0$ and $1$ every $10$ s.
\end{enumerate}
Their simulation results are plotted in Fig. \ref{fig_twotank}, \ref{fig_bilinear}, \ref{fig_CSTR} and \ref{fig_VanderPol}, respectively. The 
\textit{Two tank} and \textit{Bilinear motor} problem have no unknown disturbances and the NMPC, SL-MPC and our IA-MPC method generate the same offset-free tracking performances subject to the input constraints in their noise-free scenario, which are plotted in Fig. \ref{fig_twotank_a} and \ref{fig_bilinear_a}. The \textit{CSTR} and \textit{Van der Pol oscillator} problems have the unknown time-varying disturbance, it's found that our IA-MPC method generate the better offset-free tracking performances than the NMPC and SL-MPC method under the given EKF setting in their noise-free scenario, which are plotted in Fig. \ref{fig_CSTR_a} and \ref{fig_VanderPol_a}. Under the process noises, our IA-MPC method generate the better noise robustness than the NMPC and SL-MPC method in the four problems, which are plotted in Fig. \ref{fig_twotank_b}, \ref{fig_bilinear_b}, \ref{fig_CSTR_b} and \ref{fig_VanderPol_b}. As expected, the online computation time of the exact NMPC method is much longer than the approximate SL-MPC and IA-MPC method, and the online computation time of the SL-MPC and IA-MPC method are almost the same since they both utilized our developed construction-free CDAL method, having no online construction time.

\section{Conclusion}
This paper proposed a novel interpretative and adaptive MPC (IA-MPC) method for nonlinear systems, which was inspired by the SL-MPC with EKF method. In our IA-MPC method, a linear state-space model is firstly obtained by performing the linearization of a first-principle-based model at the initial point, and then an equivalent ARX model is obtained via the SS-to-ARX transformation. This novel initialization of ARX model allows us to keep the interpretability of a first-principle-based model and the adaptivity of the ARX model. The closed-loop control involves the EKF algorithm to update the ARX model parameters recursively and our previously developed construction-free CDAL-ARX algorithm to calculate the MPC control input. Note that our proposed implementation of our IA-MPC method is well suited for embedded platforms, thanks to its library-free C-code simple enough. The effectiveness of our IA-MPC method was illustrated by four nonlinear typical benchmarks. 

Predictably our IA-MPC method has advantages over SL-MPC for high state-dimensional first-principle-based plants, such as distributed parameter models or computational fluid dynamic models, which are our main future applications.

The SL-MPC with EKF method lacks rigorous theoretical guarantee but is still widely adopted in many practical applications for its effectiveness. Its connected IA-MPC method presented in this paper, is also practically useful. Our future work will also focus on analyzing the convergence guarantee of the IA-MPC method, borrowing from work such as adaptive control of linear time-varying systems, or exploiting the covariance matrix that directly quantifies the uncertainty associated with ARX model parameters to setup robust MPC schemes.

\bibliographystyle{unsrt}
\bibliography{ref} 
\end{document}